\documentclass[aps,prl,reprint,groupedaddress]{revtex4-1}

\usepackage[colorlinks=true,linkcolor=red]{hyperref}
\usepackage{amsmath}
\usepackage{amsfonts}
\usepackage{amssymb}
\usepackage{verbatim}
\usepackage{times}

\newcommand{\ket}[1]{\left| #1\right\rangle}
\newcommand{\bra}[1]{\left\langle #1\right|}
\newcommand{\braket}[2]{\left\langle #1| #2\right\rangle}
\newcommand{\Ord}[1]{{O}\left(#1\right)}

\newcommand{\avg}[1]{\left< #1 \right>}
\newcommand{\rp}{\overline{P}}

\def\cN{\mathcal N}

\def\dis{h}
\newcommand{\IPR}[1]{\mathrm{IPR}_{#1}}

\usepackage{graphicx}
\graphicspath{{../}{../pictures}}

\begin{document}


\title{Ergodicity breaking in a model showing many-body localization}



\author{A.De Luca}
\affiliation{SISSA - 
via Bonomea 265, 34136, Trieste, Italy}
\affiliation{INFN, Sezione di Trieste -
via Valerio 2, 34127 Trieste, Italy
}
\author{A.Scardicchio}
\affiliation{Abdus Salam ICTP - 
Strada Costiera 11, 34151, Trieste, Italy}
\affiliation{INFN, Sezione di Trieste -
via Valerio 2, 34127 Trieste, Italy
}

\begin{abstract}
We study the breaking of ergodicity measured in terms of return probability in the evolution of a quantum state of a spin chain. In the non ergodic phase a quantum state evolves in a much smaller fraction of the Hilbert space than would be allowed by the conservation of extensive observables. By the anomalous scaling of the participation ratios with system size we are led to consider the distribution of the wave function coefficients, a standard observable in modern studies of Anderson localization. We finally present a criterion for the identification of the ergodicity breaking (many-body localization) transition based on these distributions which is quite robust and well suited for numerical investigations of a broad class of problems.
\end{abstract}

\maketitle

The question of whether an Anderson localization (AL) transition \cite{anderson1958absence} can occur in a system of interacting particles has been recently suggested to have a positive answer \cite{basko2006problem,basko2006metal}.
The mechanism which underpins this effect (dubbed many-body localization
or MBL transition) requires the interaction to act in a substantially non-perturbative way, therefore providing
an example of how disorder and strong interactions interplay in a quantum theory. 

The natural setup to study the MBL transition is the dynamics (these were also the terms of the question posed in \cite{anderson1958absence}) and in this perspective it is a question about the foundations of statistical mechanics, namely, on the validity of the ergodic hypothesis. MBL also presents the terms in which a quantum glass can be defined and from there it is only a small leap to conjecturing that MBL is a natural ingredient for hard computational \emph{quantum} problems \cite{altshuler2009anderson,young2010first} (as Ising spin glasses are a natural scenario to discuss the Physics of hard combinatorial optimization problems \cite{hartmann2002optimization,nishimori2001statistical}). Even by neglecting the implications for experiments (and there are many \cite{basko2007experimental, aleiner2010finitetemperature}) the topic should be considered worth of serious investigation.

One dimensional systems are particularly suited for studying MBL because the single particle
spectrum is completely localized for arbitrarily small disorder, and
therefore any observation of delocalization must be attributed to the interaction.
In this paper, we analyze in detail the ergodicity properties of an XXZ chain with random fields. 
This particular example has already provided different indications of
the MBL transition for sufficiently large disorder: in \cite{pal2010mb}
correlation functions and spectral properties were studied, while in
\cite{znidaric2008many,Bardason2012} tDMRG was used to investigate the
different saturation properties of the entanglement entropy in the two phases.
While the existence of a transition in the dynamics of this model is now almost
certain, its precise location, the possible existence of a critical phase and
the nature of the phases that it separates are subject of debate. This should
not be regarded as  a debate about a particular spin chain but rather as an
attempt at characterizing as much as possible the differences between MBL and
AL.

Consider the real time evolution of a state $\ket{\psi_0}$ as it is encoded into the Green's function ($\hbar=1$ in the rest of the paper)
\begin{equation}
 \label{greenfun}
G(t) \equiv \bra{\psi_0}e^{-itH}\ket{\psi_0} \; .
\end{equation}
We introduce the inverse participation ratios as the moments
\begin{equation}
 \label{prq}
\IPR{q} = \sum_{E} |\braket{E}{\psi_0}|^{2q}.
\end{equation}
where the sum runs over the full set of eigenstates $\ket{E}$. The long time average of the survival probability (the
average removes some finite size effects like quasi-periodicity etc.) can be
expressed as
\begin{equation}
\rp \equiv \lim_{\tau\to\infty}\frac{1}{\tau}\int_0^\tau dt\, \left| G(t)\right|^2 =
\IPR{2}\; .
\end{equation}
Here $(\IPR 2)^{-1}$ is therefore a measure of the portion of explored Hilbert
space during the quantum dynamics and it is usually dubbed \textit{
participation ratio} (PR). Analogously, higher order $\IPR{q}$'s describe finer details of the dynamics.

Let us now comment on the choice of a suitable initial state for a
\emph{Gedankenexperiment} aimed at testing the breaking of ergodicity. First of
all, consider what happens if we take a random state in the Hilbert space
(therefore not an eigenstate) conditioned just to have an expectation value of
the energy $E$ (with high probability, for a
random state and a local Hamiltonian the standard deviation $\delta=\Ord{N^{1/2}}\ll E= \Ord{N}$). The average values of local operators in this state will not show signs of ergodicity breaking. In
fact, even at very large disorder there are states very close in
energy ($\Delta E=\Ord{e^{-S}}$, where $S$ is the microcanonical entropy at
energy $E$) which are macroscopically different and the expectation value of a
local operator will be the average of its values in these localized
eigenstates, concealing the effect of disorder (as expected from the ergodic
theorem \cite{vonneumann1929beweis}). If we want to observe the effect of
disorder on the dynamics, 
a reasonable prescription consists in choosing an eigenstate of the part of the Hamiltonian which dominates in the strong disorder limit. Starting the dynamics coincides then with \emph{turning on} the rest of the Hamiltonian.
In the delocalized phase, during the quantum dynamics, the motion covers a
finite fraction of the full Hilbert space (each eigenstate being individually
thermal, the so-called ``eigenstate thermalization hypothesis'' (ETH)
\cite{srednicki1994chaos, srednicki1999thermal,rigol2012thermalization}). 
Instead, in presence of strong disorder, ergodicity breaks down and the many-body wave
function motion is constrained on a small section of the full Hilbert space. 

We also believe that this point of view on MBL is what better brings forward its implications for quantum computation (or at least for the performance of the Adiabatic Algorithm \cite{farhi2001quantum}). In the localized phase the system gets frozen, the dynamics unable to efficiently explore the Hilbert space, so the algorithm is not efficient in finding the ground state \cite{farhi2001quantum,altshuler2009anderson,young2010first}. 

This view on the MBL transition will be the focus of this paper. We will show how the usual criteria for detecting AL need to be tweaked to capture the MBL transition; we will study the IPR's and will show how, although much information is contained in them, it is actually necessary to study the distribution of wave-function coefficients $\braket{\psi_0}{E}$, which is heavily tailed both in the localized and delocalized regions. 

We consider the Hamiltonian
\begin{equation}
H=-J\sum_{i=1}^N(s^x_i s^x_{i+1}+s^y_i s^y_{i+1})-\Delta\sum_{i=1}^N s^z_i
s^z_{i+1}
-\sum_{i=1}^Nh_i s^z_i,
\label{eq:Ham}
\end{equation}
with periodic boundary conditions. As the Hamiltonian
commutes with the total $z$ spin $S^z=\sum_i s_i^z$, we focus on the subspace with $S^z=0$. 
The random fields are chosen from a box distribution
$h_i\in[-\dis,\dis]$. The model can be cast into
a theory of fermions ($S^z = 0$ corresponds to half-filling), with on-site
disorder $h_i$. 

The $\Delta s^z s^z$ term can be written as a two-body, point-like interaction for the fermions and for
zero temperature it can be included perturbatively or non-perturbatively
\cite{giamarchi1988anderson} leading to an interesting phase diagram. When $\Delta=0$ the fermions are free, an arbitrarily small disorder localizes the entire spectrum and therefore ergodicity is broken for any $h>h_c=0$. As $\Delta$ is increased MBL would appear as a peculiar phase transition (possibly even at infinite temperature) at a critical $h_c$ increasing away from zero. 
On the other hand, for $\Delta\gg J$ the disorder necessary to break ergodicity should \emph{decrease} again. In fact, for large $\Delta$ the relevant degrees of freedom are the domain walls of the classical Ising chain obtained by setting $J=0$ in (\ref{eq:Ham}). Longer domain walls have smaller hopping matrix elements and therefore they are more prone to localization than the fermions at $J\gg \Delta$. Once a few of these large domain walls have frozen, ergodicity can be considered broken and this occurs for smaller $h$, since both the effective hopping and interaction are smaller (effective randomness is always $h$). Here we present results of exact diagonalization for $\Delta=J=1$, where the delocalized phase is largest.

\section{Return probability}
According to the discussion of the previous section, we should test ergodicity by taking an initial state $\psi_0$ as one of the  $\cN=\binom{N}{N/2}$ configuration of spins $\ket{a}$ polarized along the $z$ or $-z$ direction, (e.g.\ $\ket{a}=\ket{\uparrow\downarrow...}$).
We need to stress a major difference in the  behavior of $\IPR{2}$ in the localized and delocalized phases between AL and the present situation.
While in the former one can distinguish the two phases by the participation ratio being
$\Ord{1}$ or not in the thermodynamic limit, this is not a
sufficient criterion for us. For a many-body state, even in absence of interaction, $\IPR{2}$ will be exponentially small in $N$ also in presence of strong disorder, simply because each degree of freedom will have a localization length small but finite, corresponding to an individual \textit{participation ratio} smaller than $1$: multiplication of $\Ord{N}$ of these factors leads to an exponentially small $\IPR{2}$. We need to correct the previous criterion by requiring that the delocalized and localized phase are distinguished by whether the ratio $\IPR{2}/\cN^{-1}$ is $\Ord{1}$ or not. The other $\IPR{q}$'s, properly rescaled with powers of the Hilbert space dimension $\cN$, also represent indicators of ergodicity breaking. 

\begin{figure}
 \includegraphics[width=1\columnwidth]{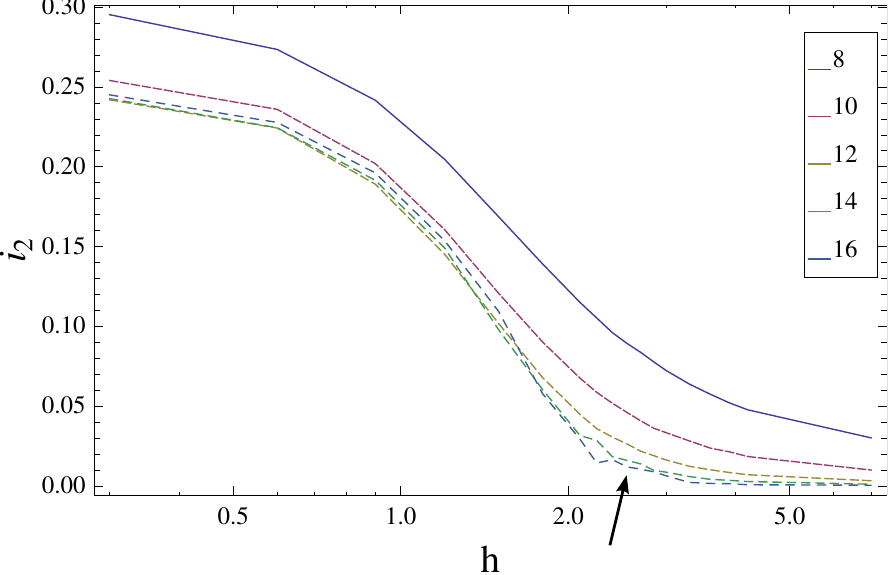}
 \caption{Average fraction of occupied Hilbert space as a function of h for different system
sizes $N = 8$ to $16$ using exact diagonalization. Notice how the limit for $h\to 0$ is different from $1/3$ which is the RMT prediction.}
\label{fractionHilbert}
\end{figure}

However, as far as averages over the initial states are involved we have found that PR's have better finite-size behaviors (more on this later), so we considered:
\begin{equation}
 \label{intensiveIPR}
I_q^{(N)}(\dis) \equiv \left\langle \frac{\IPR{q}^{-1}}{\cN^{q-1}}
\right\rangle_{\{\dis\}, a} \; .
\end{equation}
where the subscripts in the average correspond to disorder realizations
(indicated with $h$) and initial spin configuration $\ket{a}$\footnote{The number of realizations goes from 10000 for small sizes till about 100 for the maximum size $N = 16$}.
In particular the data for $I_2$, shown in Fig.~\ref{fractionHilbert}, are consistent with the $\lim_{N\to \infty}I_2^{(N)}(h)=i_2(h)$ where $i_2(h) = 0$, for $h>h_c=2.7\pm0.3$, although the finite-size corrections are strong already at $h\gtrsim1.5$. 
The prediction of $1/3$ for $i_2$ at small $h$ coming from the GOE ensemble, although qualitatively correct, is quantitative inaccurate. This could be however a consequence of the many-body structure in the finite size scaling that we did not take into account up to now.

\begin{figure}
\includegraphics[width=0.5\textwidth]{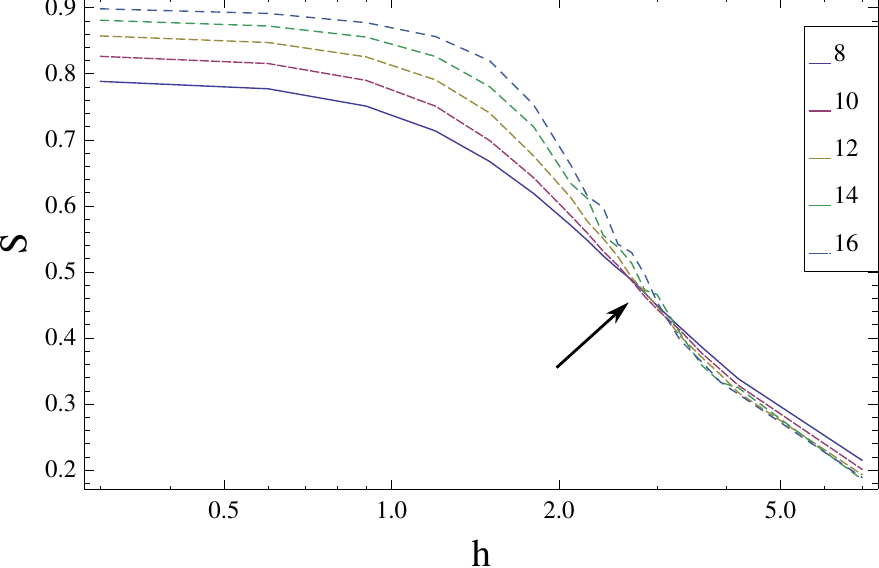}
\caption{Average diagonal entropy as a function of disorder strength for different sizes $N =
8,10,12,14,16$. From the $N$-dependence the transition is identified at $h_c\simeq 2.7\pm 0.3$.}
\label{entropySize}
\end{figure}

A similar information is obtained by the diagonal entropy
\begin{equation}
S^{(N)}=\lim_{q\to 1}\frac{\avg{\IPR{q-1}}}{(q-1)\ln\cN},
\label{eq:diagentr}
\end{equation}
which is plotted for varying $h$ in Figure \ref{entropySize} and also this quantity is clearly far from its thermodynamic limit of $S=1$ in the delocalized phase. If we identify the critical point (see the arrows in Fig.~\ref{fractionHilbert} and Fig.~\ref{entropySize}) as the place where the $N$ dependence sets in (for $I_2$) or drops out (for $S$) then both quantities identify a critical point consistent with $h_c=2.7\pm 0.3$ consistently with the findings of \cite{pal2010mb}.

The diagonal entropy and the IPR's show that the wave function covers a number of sites that grows exponentially with the system size, although the exponent is smaller than in the ergodic phase.
This suggests that in a many-body system, the localized phase is necessarily characterized by the breaking
of ergodicity, but not necessarily by a concrete localization ($\IPR{2} \simeq O(1)$). However, to pinpoint the transition and understand the reasons of this anomalous scalings we should analyze the full probability distribution of $|\braket{a}{E}|^2$.

\section{Distribution of wave function amplitudes}
If one considers the various $\IPR{q}$ averaged over $\ket{a}$, one observes a peculiar scaling with $N$ of each of them, which can be considered as due to large fractal dimensions. In this scenario, the safest observable to consider is the distribution of the properly rescaled wave function coefficients.  As we are
interested in typical states (infinite temperature) we will not follow the usual route of fixing
the energy of the state but we will rather integrate over the whole spectrum. In
the thermodynamic limit this corresponds to energy density $E/N=0$.
We will consider therefore the average over eigenstates, initial states and disorder realizations: 
\begin{equation}
\label{distrib}
\phi(x,N)=\avg{\delta(x-\cN |\braket{a}{E}|^2)}_{a,E,\{h\}}.
\end{equation}
In the following we will drop the subscripts in the averages. This function depends both on $x$ and $N$ in general but in the ergodic delocalized phase, as $\cN$ plays the role of the space volume, we see that the dependence on $N$ drops out \cite{mirlin1993statistics, mirlin1994statistical, mirlin1994distribution}.

We can then write the various IPR's as
\begin{equation}
\label{prqdistrib}
\avg{\IPR{q}} = \cN^{1-q}\int_0^\infty dx\ x^q\phi(x).
\end{equation}

\begin{figure}[htbp]
\begin{center}
\includegraphics[width=0.95\columnwidth]{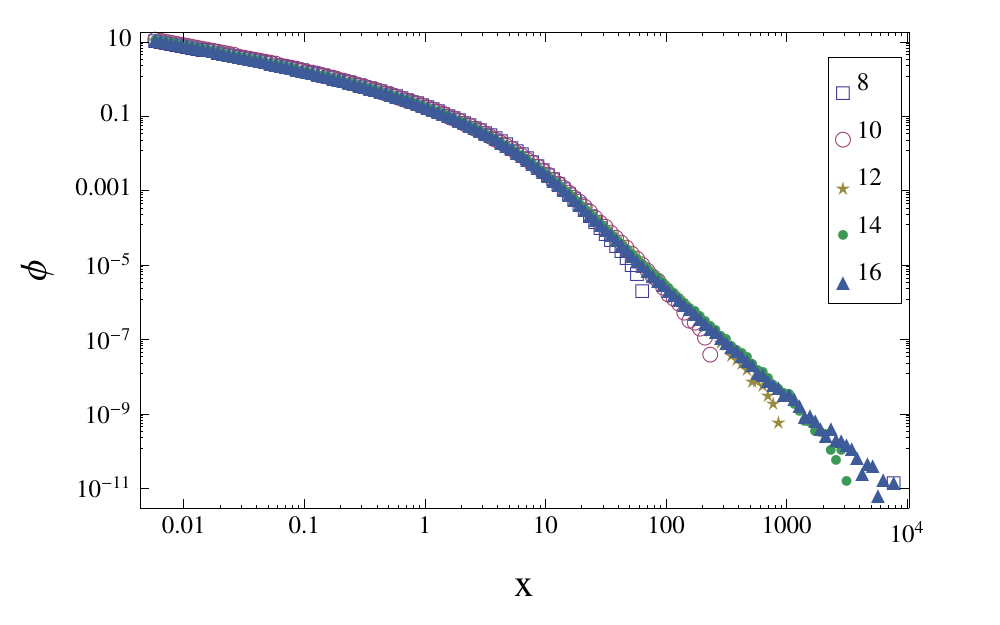}
\includegraphics[width=0.95\columnwidth]{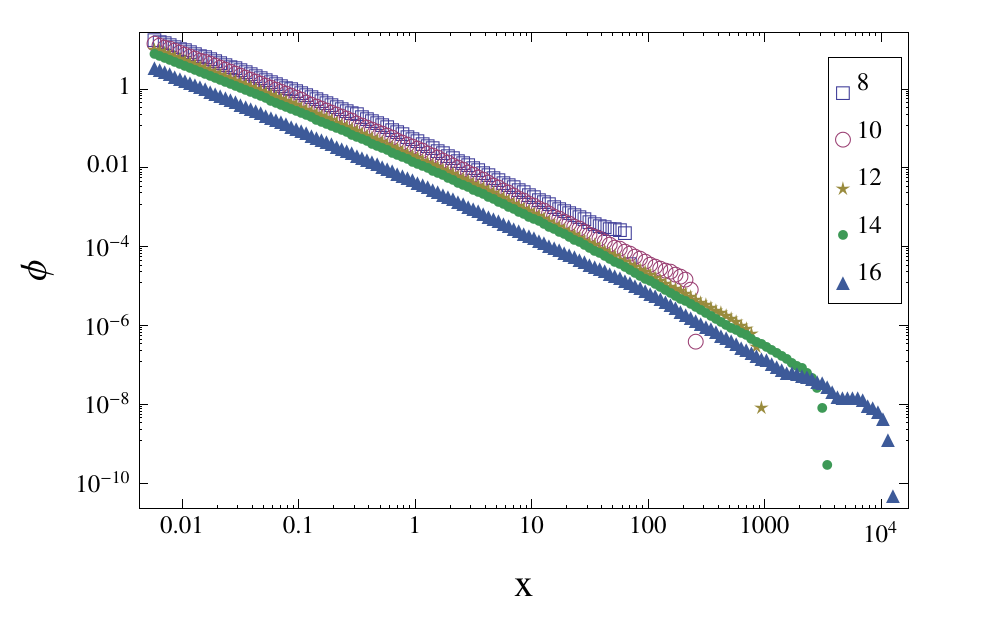}
\caption{The distribution $\phi$ of scaled wave function amplitudes $x=\cN |\braket{a}{E}|^2$
for different values of $h$. Upper panel: $h=1.2$ in the middle of the ergodic phase where the scaling is perfectly verified, lower panel $h=4.2$ in the many-body localized phase. In each figure the different curves correspond to different values of
$N$, from 8 to 16. Each curve is obtained by binning of not less than $3\ 10^6$
squared amplitudes.}
\label{fig:phi-h}
\end{center}
\end{figure}

Illustrative plots are shown for different regimes in Fig.~\ref{fig:phi-h}.
As we said, even though in the ergodic phase, with this scaling the curves for
different sizes collapse (similarly to AL), the distribution has an elbow at $x\sim 1$ and we find
\begin{equation}
\label{delocFit}
\phi(x)\propto
\begin{cases}
x^{-\alpha} & \text{if } x \lesssim 1 \\
x^{-\beta}  & \text{if } x \gtrsim 10,
\end{cases}
\end{equation}
where $\alpha,\beta$ depend on $h$. We have $\alpha<1<\beta$ ensuring the normalization of the distribution function in the delocalized phase and their values are almost independent of $N$ for the largest sizes explored.\footnote{A residual $N$ dependence is found in the left tails, at $x\ll 10^{-3}$, that part of the distribution reaching its asymptotic form for larger $N$ ($N\geq 14$).} This is an uncommon distribution for the quantity $x$: in the Anderson model usually $\alpha=1/2$ and the large $x$ behavior is exponential \cite{mirlin2000statistics} reminiscent of the Porter-Thomas distribution of RMT \cite{porter1965statistical}. Comparing the power-law tail with the exponential one of the delocalized phase in the Anderson problem, we conclude that already deep in the delocalized region, there are sign of pre-localization.
The almost perfect collapse of the curves in the upper panel of Fig.~\ref{fig:phi-h} allows a much better finite size scaling analysis than any of its moments.

\begin{figure}
\includegraphics[width=0.9\columnwidth]{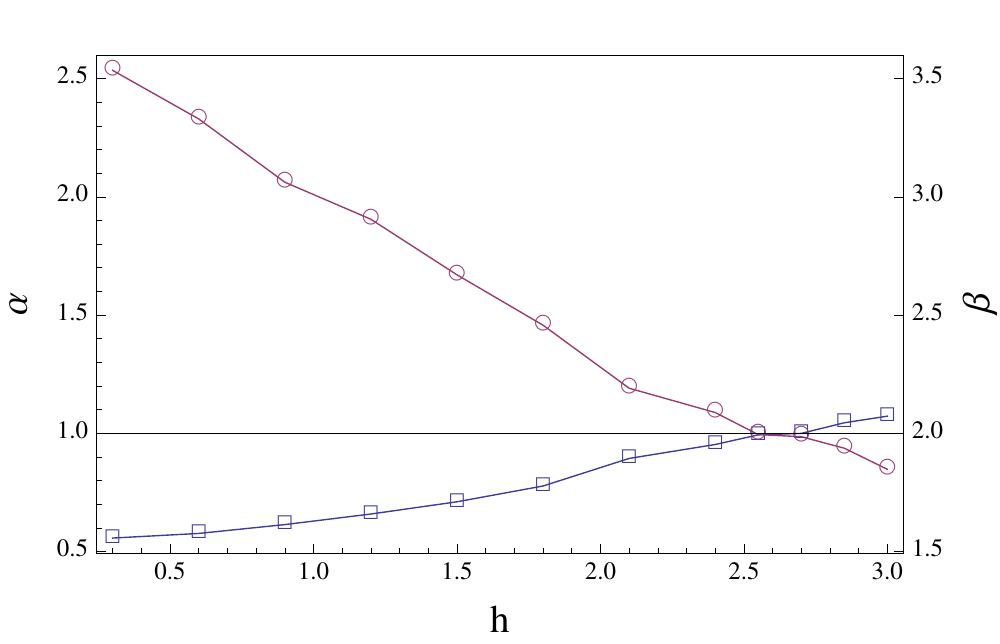}
\caption{The value of the exponent $\alpha$ (blue squares) and $\beta$ (pink circles) in Eq.\ (\ref{delocFit}) for $N=16$ (these exponents are independent of $N$ within the symbol size). The exponent $\alpha$ crosses the value $1$ required by summability, which occurs at $h\simeq 2.55\pm 0.05$, precisely where (within errors) $\beta$ crosses the value $2$, required for the existence of the first moment (normalization of the wave function).}
\label{fig:alphabeta}
\end{figure}

As $h$ approaches $h_c\simeq 2.6$ the elbow smoothens and $\alpha\to 1$ so that we can identify $h_c$ as the point at which $\alpha=1$, the distribution stops being summable and necessarily the independence on $N$ ceases.\footnote{As $\avg{x}=1$ is fixed by normalization the divergence of $\avg{1}$ implies a divergence of the first moment as well. In fact, $\beta=2$ occurs at the same value of $h_c$.} This occurs at $h_c=2.55\pm0.05$ as it can be seen in Fig.~\ref{fig:alphabeta}. An explicit $N$-dependence of $\phi$ means that the scaling of all the IPR's and of the diagonal entropy with $N$ change abruptly and ergodicity is broken.

The exponent $\beta$ governs the scaling of the various $\IPR{q}$'s. For $0<q<\beta-1$ the integral in \eqref{prqdistrib} is finite and $\IPR{q}\sim \cN^{1-q}$. If instead $\beta-1<q$, since the integral in \eqref{prqdistrib} is divergent
the average of $\IPR{q}$ over the initial states $\ket{a}$ does not exist, but the typical value for a state should be found by looking at the sum of $\cN$ independent and identically distributed variables $x_a^q$. One then finds the probability density for $\sum_{a\leq \cN}x_a^q\equiv Y$ (by computing and then inverting its Laplace transform, provided $\beta>2$) as
\begin{equation}
\label{typicalPq}
P(Y)\propto Y^{-\frac{3-\gamma}{4-2\gamma}}\exp\left(-C\left(\frac{\cN^{\frac{1}{\gamma-1}}}{Y}\right)^{\frac{\gamma-1}{2-\gamma}}\right),
\end{equation}
where $\gamma=1+(\beta-1)/q$, ($1<\gamma<2$) and $C$ is a constant of $\Ord{1}$. This distribution has a power law tail but the typical value of the sum is set by the exponential as $Y\sim \cN^{1/(\gamma-1)}\gg \cN$. This implies typical values of the $\IPR{q}$ of a state, when $q>\beta-1$:
\begin{equation}
\IPR{q}^{(N)}\sim\cN^{-q+\frac{q}{\beta-1}}.
\end{equation}
The different $\IPR{q}$ define different ``critical points" $h_q$ solutions of $\beta(h_q)=q+1$. The real transition, signaled by an explicit $N$-dependence of full distribution $\phi$ can then be identified by the diagonal entropy (\ref{eq:diagentr}), or the limit as $q\to 1$ of $\IPR{q}$, therefore when $\beta=2$. What is the possible origin of the power-law tail at large $x$?\footnote{We thank V.Oghanesyan for discussions on this point.} This can be linked with the existence of a many-body mobility edge at some energy $E^*(h)$, where eigenstates occupy $\Ord{\cN}$ sites above $E^*$ and $\Ord{\cN^a}$ ($a(h)<1$) below $E^*$ and to a competition between the canonical entropy (the logarithm of the number of states between energy $E$ and $E+dE$) and the diagonal entropy multiplied by $q$. This phenomenon deserves better investigation in a future work.

Summarizing, the coincident divergence of $\avg{1}$ (a non-summability of $\phi(x)$ at small $x$), and of $\avg{x}$ (non summability of $x\phi(x)$ at large $x$) signal the beginning of the localized region. This implies an accumulation of wave-function
amplitudes towards small values typical of localized states\cite{mirlin1994distribution}.
We expect then that the scaling of the \emph{typical coefficients} changes abruptly at the onset of the region in which ergodicity is broken but the wave functions are still extended.

This suggests a description of the localized phase in which a typical eigenstate is described by negligible weight on ample regions of the Hilbert space, which is reminiscent of the ``small branching number" Bethe lattice picture of \cite{altshuler1997quasiparticle,basko2006problem} and of the eigenstates of a disordered but integrable model \cite{buccheri2011structure}.

\section{Similarities with AL on the Bethe lattice \cite{abou1973selfconsistent,mirlin1994distribution}}  Our case shows three differences from this classic topic: {\it 1)} our lattice has connectivity $\Ord{N}\gg\Ord{1}$ (but still $\ll\Ord{\cN}$, the volume of the system), {\it 2)} the on-site disorder potentials of neighboring configurations $a$ and $b$ are strongly correlated ($E_a-E_b=h_{i+1}-h_i\ll E_{a},E_{b}$) and {\it 3)} our lattice is not random at all. In order to identify which of these three ingredients are necessary to preserve this phenomenology of the distribution functions we have investigated numerically a random graph with $\cN$ nodes and fixed connectivity $N/2$ and \emph{independent} random energies $\epsilon_i$ on each node. We observe the same qualitative features in the distribution $\phi(x)$, even for small $\dis$. On the contrary, for the Anderson model on a Bethe lattice with connectivity $\Ord{1}$ in the ergodic region we observe an exponential (or possibly stretched-exponential) 
tail at large $x$ (the data will be presented in a future publication). Therefore we conjecture that the necessary requirement for the large $x$ power-law tail is the growing connectivity, and that one can get rid of the correlation of the energies and the specific topology of the hypercube. 

This confirms that we have the right to look at MBL as a localization phenomenon on a Bethe lattice with asymptotically large connectivity, a problem amenable of analytic treatment, beyond the locator expansion \cite{altshuler1997quasiparticle,abou1973selfconsistent}.

\section{Summary and conclusions}
We have investigated the behavior of the return (or survival) probability as a possible detector of ergodicity breaking and of the MBL transition. We have shown how this question leads to the necessity of a thorough study of the distribution of the wave-function amplitudes of the eigenstates averaged over all energy spectrum.\footnote{We emphasize once more that our study has been limited to the distribution of wave-function coefficients averaged over all the spectrum because of its connection to the question of breaking of ergodicity. Distributions at a fixed energy will probably have different behaviors and will be subject of future studies.} We then identified the major changes which occur to this distribution at the MBL transition point. The delocalized, ergodic phase looks more localized than the corresponding single-particle AL and RMT does not seem to be a good approximation, not even deep in the delocalized region.     
The localized region seems very akin to the case of single particle AL on the Bethe 
lattice with connectivity $\Ord{1}$, in particular the distribution functions of the amplitudes show a small-$x$ accumulation which points towards wave functions localized in configuration space. We have also identified similarities and differences with this better studied case and suggested what are the necessary ingredients for a viable analytical study of MBL. 

We gratefully acknowledge many useful and sometimes enlightening discussions with D.Huse, V.Kravtsov and M.Muller. We thank the Galileo Galilei Institute for Theoretical Physics for the hospitality and the INFN for partial support during the completion of this work.

\bibliography{LocXXZBib}
\end{document}